\documentclass[usenatbib]{mn2e}
\usepackage{ifthen}
\newcommand{\mr}[1]{\mathrm{#1}}

\newcommand{\m}{$^{-1}$}

\ifthenelse{\isundefined{\nonaas}}{}{

\bibpunct{(}{)}{;}{a}{}{,}

  \renewenvironment{thebibliography}[1]{%
    \begin{oldthebibliography}{#1}%
      \setlength{\parskip}{0ex}%
      \parindent 0ex%
      \setlength{\itemsep}{0ex}%
  }%
  {%
    \end{oldthebibliography}%
  }

\def\araa{{ARA\&A}}		
\def\apj{{ApJ}}			
\def\apjl{{ApJ}}		
\def\apjs{{ApJS}}		
\def\ao{{Appl.~Opt.}}		
\def\aap{{A\&A}}		
\def\mnras{{MNRAS}}		
\def\prd{{Phys.~Rev.~D}}	

}

\usepackage{ifpdf}

\ifpdf
\usepackage[pdftex]{graphicx}
\usepackage[update,prepend,verbose]{epstopdf}
\else
\usepackage{graphicx}
\fi

\newcommand{\mode}{submit}

\begin{document}

\title[$M_X/M_L$ relation]{Evidence for Non-Hydrostatic Gas from the Cluster X-ray to Lensing Mass Ratio}
\author[A. Mahdavi et al.]{A. Mahdavi$^1$, H. Hoekstra$^{1,2}$, A. Babul$^1$, J. P. Henry$^3$
\\
$^1$University of Victoria, Elliott Building, 3800 Finnerty Road, Victoria, BC V8P 5C2 Canada\\
$^2$Alfred P. Sloan Research Fellow\\
$^3$Institute for Astronomy, University of Hawaii, 2680 Woodlawn Drive, Honolulu, HI 96822}
\maketitle

\newcommand{\rata}{1.03 \pm 0.07}
\newcommand{\ratb}{0.90 \pm 0.09}
\newcommand{\ratc}{0.78 \pm 0.09}
\newcommand{\twrata}{1.06 \pm 0.07}
\newcommand{\twratb}{0.96 \pm 0.09}
\newcommand{\twratc}{0.85 \pm 0.10}

\begin{abstract}

Using a uniform analysis procedure, we measure spatially resolved weak
gravitational lensing and hydrostatic X-ray masses for a sample of 18
clusters of galaxies.  We find a radial trend in the X-ray to lensing
mass ratio: at $r_{2500}$ we obtain a ratio $M_X/M_L = \rata$ which
decreases to $M_X/M_L = \ratc$ at $r_{500}$. This difference is
significant at $3\sigma$ once we account for correlations between the
measurements.  We show that correcting the lensing mass for excess
correlated structure outside the virial radius slightly reduces, but
does not eliminate this trend. An X-ray mass underestimate, perhaps
due to nonthermal pressure support, can explain the residual
trend. The trend is not correlated with the presence or absence of a
cool core. We also examine the cluster gas fraction $f_\mr{gas}$ and
find no correlation with $M_L$, an important result for techniques
that aim to determine cosmological parameters using $f_\mr{gas}$.

\end{abstract}

\begin{keywords}
Gravitational lensing - X-rays: galaxies: clusters - dark matter
- galaxies: clusters: general
\end{keywords}

\section{Introduction}

Rich clusters of galaxies host the most massive collapsed dark matter
halos. The X-ray emitting intracluster medium (ICM) bound to this halo
can be a useful tracer of the dark matter content of the
cluster. Under the assumption of hydrostatic equilibrium the gradients
of the gas pressure $P$ and total gravitational potential $\Phi$ are
related by the simple differential equation
\begin{equation}
\nabla P = \rho_g \nabla \Phi
\end{equation}
where $\rho_g$ is the gas density. The hydrostatic method is in
principle powerful, because spatially resolved measurements of the gas
pressure can constrain the shape of the halo and hence yield useful
limits on fundamental dark matter properties and cosmology. 

In reality, we do not know the reliability of hydrostatic mass
estimates. Not only are merging clusters of galaxies in a
nonhydrostatic state, but the plasma in many apparently relaxed
systems may be affected by additional nonequilibrium processes, which
serve to boost $P$ and hence cause an underestimate of the cluster
mass from X-ray observations of the thermal bremsstrahlung emission.
In early hydrodynamic N-body simulations, \cite{Evrard90} found the
first signs of this underestimate, attributing it to incomplete
thermalization of the ICM. More recent N-body work has shown that
energy input from active galactic nuclei, pressure support from
turbulence and residual bulk motions, and variations in the merging
history may contribute substantially to this systematic bias
\citep{Dolag05,Faltenbacher05,Rasia06,Nagai07,Ameglio07}. Importantly,
constraints on dark energy from the cluster mass function are highly
sensitive to intrinsic scatter and systematic error in the
mass-observable relation \citep{Lima05}; the scatter also
significantly influences the normalization of the primordial
fluctuation spectrum $\sigma_8$ \citep{Balogh06}.

Comparison with independent methods can provide a powerful means of
checking the reliability of X-ray mass measurements. Here we focus on
comparison of hydrostatic X-ray masses with those derived from weak
gravitational lensing. Unlike the hydrostatic method, gravitational
lensing does not require assumptions regarding the dynamical state of
the cluster.  Early work comparing lensing and X-ray masses showed
that while strong gravitational lensing masses in merging clusters
sometimes exceeded X-ray masses by factors of $\approx 1.5-2$, weak
lensing masses measured at larger radii are consistent with the X-ray
data to within $\approx 20\%$
\citep{MiraldaEscude95,Squires96b,Allen96,Squires97,Allen98b}.  This
picture is supported by subsequent ground- and space-based
observations of relaxed and merging clusters, albeit with fairly large
uncertainties
\citep{Allen02,Smith02,Hoekstra02,Ettori03,Cypriano04,Hicks06}.

Only recently has data with the accuracy required to carry out a
systematic comparison of larger samples of hydrostatic and lensing
masses become available. Recent studies, however, are apparently in
conflict. For a sample of 30 clusters, \cite{Pedersen07} report a 30\%
excess in the normalization of lensing mass-temperature (M-T) relative
to the X-ray value within\footnote{For a cluster at redshift $z$, the
overdensity radius $r_\Delta$ is the radius within which the mean
matter density is $\Delta$ times the critical density of the
universe at redshift $z$. $M_\Delta$ is the mass within that
radius.}  $r_{500}$. However, \cite{Hoekstra07} using 20 clusters
which also form the basis of this paper, report no significant excess
in the M-T normalization at $r_{2500}$.

To study the difference in X-ray and lensing masses in greater detail,
we present a sample of 18 clusters---the largest yet with accurate
spatially resolved hydrostatic as well as weak lensing masses. In our
study, we take particular care to account for possible systematic
effects in the lensing and X-ray measurement process. Our data support
a picture in which the lensing excess is negligible at $r_{2500}$, but
increases at at larger radii. In Section \ref{sec:data} we discuss our
data reduction procedure; in Section \ref{sec:trend} we discuss the
trend in the $M_X - M_L$ relation; in Section \ref{sec:bias} we
account for potential biases; in Section \ref{sec:fgas} we consider
trends in the gas fraction; and in Section \ref{sec:conclusion} we
conclude. We assume $H_0=70$ km s\m\ Mpc\m, $\Omega_M = 0.3$, and
$\Omega_\Lambda = 0.7$.

\section{Data}
\label{sec:data}

\subsection{Lensing data}

The clusters in our sample were drawn from \cite{Hoekstra07}, which
contains a weak lensing analysis of CFH12k data from the
Canada-France-Hawaii Telescope. We refer interested readers to
\cite{Hoekstra07} for details of the data reduction and weak lensing
analysis procedure. The shear measurements discussed in this paper are
identical to \cite{Hoekstra07}, but here we present updated values for
the masses. The changes are due purely to the fact that
\cite{Hoekstra07} used the Hubble Deep Field (HDF) redshift
distribution \citep{FernandezSoto99}, which is based on data taken
within a much smaller field than the \cite{Ilbert06} study, based on
the CFHT Legacy Survey data. The conversion of the lensing signal into
a physical mass estimate depends directly on the source redshift
distribution. This dependence is quantified by the parameter
$\beta_\mr{lens}=\max[0,D_{ls} /D_s]$, where $D_{ls}$ and $D_s$ are
the angular diameter distances between the lens and the source, and
the observer and the source. We note that $\beta_\mr{lens}$ is an
important parameter because it is degenerate with the projected
cluster mass.

We calculate the new $\beta_\mr{lens}$ values using the photometric
redshift distributions derived by \cite{Ilbert06}. The resulting
values are listed in Table~\ref{tbl:lensing}. The mean redshift
derived by \cite{Ilbert06} is higher \citep[also
  see][]{Benjamin07} resulting in a $\sim 10\%$ reduction in the
inferred cluster mass compared to those in \cite{Hoekstra07}. As the
\cite{Ilbert06} results are based on four independent pointings, we
can also examine the field-to-field variation in $\beta_\mr{lens}$.
We find this to be a negligible effect, with a dispersion of only a
$1-2$ percent. Although systematic biases may still be present, the
results from \cite{Ilbert06} provide a significant improvement over
previous studies. Consequently, we expect that the current uncertainty
in the source redshift distribution affects our mass estimates only at
the few percent level.

We base our lensing masses on the aperture mass estimates \citep[for details
see the discussion in \S3.5 in][]{Hoekstra07}. This approach has
the advantage that it is practically model independent. Additionally,
as the mass estimate relies only on shear measurements at large radii,
contamination by cluster members is minimal. \cite{Hoekstra07} removed
galaxies that lie on the cluster red-sequence and boosted the signal
based on excess number counts of galaxies. As an extreme scenario we
omitted those corrections and found that the lensing masses change by
only a few percent at most. Hence our masses are robust against
contamination by cluster members at the percent level.

The weak lensing signal, however, only provides a direct estimate of
the {\it projected} mass. To calculate 3D masses (such as $M_{2500}$,
$M_{1000}$, and $M_{500}$) from the model-independent 2D aperture
masses we project and renormalize a density profile of the form
$\rho_\mr{tot}(r) \propto r^{-1} (r_{200}+c r)^{-2}$ \citep{NFW}. The
relationship between the concentration $c$ and the virial mass is
fixed at $c \propto M_\mr{200}^{-0.14}/(1+z)$ from numerical
simulations \citep{Bullock01}. Hence, the deprojection itself, though
well motivated based on numerical simulations, is model dependent.

\begin{figure}
\includegraphics[width=84mm]{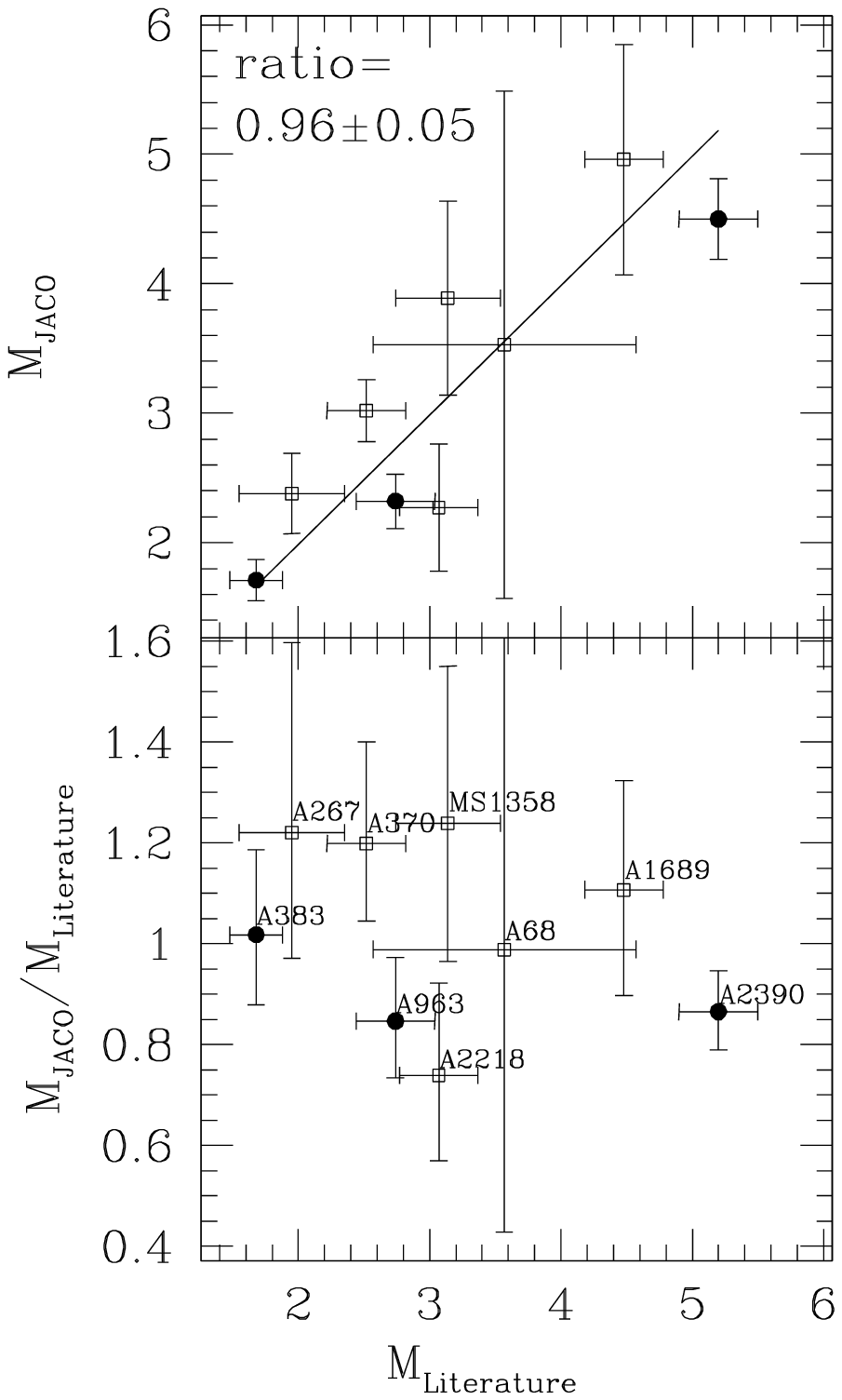}
\caption{Comparison of JACO measurements of $M_{2500}$ with other
  recent Chandra measurement for a subset of 9 clusters. All masses
  are in units of $10^{14} M_\odot$. The Abell 383, 963, and 2390
  masses (filled points) are from \protect\cite{Allen07}; the
  remainder (unfilled points) are from \protect\cite{LaRoque06}. The
  JACO masses shown are estimated using X-ray data only, rather than
  calculated using the lensing values of $r_{2500}$ as in Table
  \protect\ref{tbl:xray}.}
\label{fig:xraycomp}
\end{figure}

\begin{table*}
\begin{minipage}{126mm}
\caption{Weak-Lensing Measurements}
\label{tbl:lensing}
\begin{tabular}{lrrcrcrcr}
\hline
Cluster & $z$ & $\beta_\mr{lens}$ & $M_{2500}$ & $r_{2500}$ & $M_{1000}$ & $r_{1000}$ & $M_{500}$ & $r_{500}$ \\
\hline
    Abell   68 & 0.255& 0.504 &  $2.56\pm0.60$ &  0.52 &  $4.51\pm1.30$ &  0.85 & $ 6.64\pm2.44$ &  1.22\\
    Abell  209 & 0.206& 0.627 &  $1.89\pm0.67$ &  0.48 &  $4.40\pm1.61$ &  0.86 &  $7.14\pm1.93$ &  1.27\\ 
    Abell  267 & 0.230& 0.522 &  $2.14\pm0.39$ &  0.49 &  $3.70\pm1.10$ &  0.80 &  $6.29\pm1.93$ &  1.21\\ 
    Abell  370 & 0.375& 0.412 &  $3.86\pm0.66$ &  0.57 &  $7.79\pm1.86$ &  0.97 & $13.27\pm3.40$ &  1.47\\ 
    Abell  383 & 0.187& 0.641 &  $0.76\pm0.33$ &  0.35 &  $2.29\pm1.04$ &  0.69 &  $4.49\pm1.96$ &  1.09\\ 
    Abell  963 & 0.206& 0.592 &  $1.39\pm0.41$ &  0.43 &  $2.50\pm0.90$ &  0.71 &  $4.16\pm1.54$ &  1.06\\ 
    Abell 1689 & 0.183& 0.639 &  $4.74\pm0.77$ &  0.65 &  $9.11\pm1.31$ &  1.10 & $14.29\pm2.40$ &  1.61\\ 
    Abell 1763 & 0.223& 0.574 &  $2.93\pm0.59$ &  0.55 &  $5.51\pm1.43$ &  0.92 & $10.47\pm2.84$ &  1.43\\ 
    Abell 2218 & 0.176& 0.644 &  $2.66\pm0.64$ &  0.54 &  $4.59\pm1.10$ &  0.88 &  $6.10\pm1.64$ &  1.22\\ 
    Abell 2219 & 0.226& 0.562 &  $3.10\pm0.60$ &  0.56 &  $6.81\pm1.79$ &  0.98 & $10.27\pm2.31$ &  1.42\\ 
    Abell 2390 & 0.228& 0.586 &  $3.04\pm0.59$ &  0.55 &  $6.29\pm1.43$ &  0.96 &  $8.79\pm1.99$ &  1.35\\ 
CL 0024.0+1652 & 0.390& 0.379 &  $3.16\pm0.63$ &  0.53 &  $5.90\pm1.66$ &  0.88 &  $9.87\pm3.36$ &  1.32\\ 
MS 0015.9+1609 & 0.547& 0.267 &  $3.74\pm0.87$ &  0.53 & $11.30\pm3.07$ &  1.03 & $19.51\pm5.77$ &  1.56\\ 
MS 0906.5+1110 & 0.170& 0.674 &  $1.99\pm0.57$ &  0.49 &  $4.87\pm1.54$ &  0.90 &  $9.46\pm2.06$ &  1.41\\ 
MS 1358.1+6245 & 0.329& 0.447 &  $2.23\pm0.51$ &  0.48 &  $4.11\pm1.33$ &  0.80 &  $6.64\pm2.66$ &  1.18\\ 
MS 1455.0+2232 & 0.257& 0.564 &  $1.51\pm0.36$ &  0.43 &  $3.11\pm1.00$ &  0.75 &  $4.83\pm1.70$ &  1.09\\ 
MS 1512.4+3647 & 0.373& 0.434 &  $0.74\pm0.36$ &  0.33 &  $1.43\pm0.74$ &  0.55 &  $2.94\pm2.11$ &  0.89\\ 
MS 1621.5+2640 & 0.428& 0.368 &  $2.03\pm0.87$ &  0.45 &  $5.40\pm1.80$ &  0.84 &  $7.64\pm2.74$ &  1.19\\ 
\hline
\end{tabular}

\medskip
All masses are in units of $10^{14} M_\odot$; all radii are in units of Mpc; $z$ is the redshift
of the clusters, and $\beta_\mr{lens}$ is a measure of the source redshift distribution.
\end{minipage}
\end{table*}

\subsection{X-ray data}

Data for 18 of the 20 clusters in \cite{Hoekstra07} are available in
the Chandra X-ray Observatory (CXO) public archive. We analyze and fit
these data using the Joint Analysis of Cluster Observations (JACO)
package \citep{Mahdavi07}. The reduction procedure follows the
detailed description in \cite{Mahdavi07}, except as discussed
below. We briefly summarize the procedure here. We reprocess the raw
CXO data with CALDB 3.3, including the charge transfer inefficiency
correction for ACIS-I data. To remove the particle background, we
subtract the appropriate blank-sky observation renormalized to match
9-12 keV count rates in the most source-free region of the data.  We
examine the residual spectra in 9-12 keV energy range for each cluster
to ensure that the particle background is cleanly subtracted.

We then extract spectra in circular annuli centered on the X-ray
surface brightness peak, masking any detected non-cluster sources. The
residual diffuse astrophysical background is fit as an additive
component as described in the analysis below.

The spectra are fit with a projected 3D cluster model.  We begin with
a mass profile consisting of an NFW dark matter distribution and a gas
density of the form
\begin{equation}
\rho_g = \sum_{i=1}^{N_\beta} \rho_i (1+r^2/r_{x,i}^2)^{-3
\beta_{i}/2},
\end{equation}
where $r_{x,i}$ are the core radii and $\beta_i$ are the slopes of the
$N_\beta$ independent``$\beta$-model gas distributions. 
We assume that the gas metallicity is of the form
\begin{equation}
Z(r) = Z_0 (1+r^2/r_z^2)^{-3 \beta_Z/2}.
\end{equation}
where $Z_0$ is the central metallicity, and $r_Z$ and $\beta_Z$
describe the metallicity profile.

The 3D temperature profile is calculated self-consistently using the
equation of hydrostatic equilibrium:
\begin{equation}
T(r) = \frac{1}{\rho_g} \left( \rho_{100} T_{100} + \frac{\mu m_p}{k} \int_{r}^{r_{100}}
  \frac{G M \rho_g}{r^{\prime 2}} \, dr^\prime \right)
\label{eq:hydrosol}
\end{equation}
where $M$ is the total mass profile, $\mu$ is the mean molecular
weight, $m_p$ is the proton mass, $r_{100}$ is the radius at which the
gas distribution is truncated, $\rho_{100}$ is the gas density at the
truncation radius, and $T_{100}$ is the temperature at that radius. We
project the resulting 3D emissivity along the line of sight, absorb it
by the galactic hydrogen column as fixed by \cite{Dickey90}, convolve
the result by the instrumental response, and compare with the measured
spectra using a $\chi^2$ statistic. A standard MEKAL plasma with
variable abundance serves as the spectral model.

The fitting procedure begins with a single $\beta$ model fit
($N_\beta=1$); if this yields a poor fit, one or two $\beta$ model
components are added until the fit becomes acceptable. In two cases
(Abell 1689 and Abell 2390) we could only obtain a good fit by also
excising the central 70 kpc, as \cite{Vikhlinin06} suggest. Other
simultaneously fit parameters are the gas metallicity profile (3 free
parameters), the dark matter profile (2 free parameters), the
normalizations of the diffuse astrophysical backgrounds (3 free
parameters), and the temperature at the truncation radius $T_{100}$.
The astrophysical backgrounds are modeled as a power law with fixed
slope 1.4 and free normalization (to model unresolved background AGN),
and a unredshifted thermal plasma with free normalization and $T <
0.5$ keV (to model the local soft X-ray background).

Accounting for the covariance of all measurable parameters, proper
treatment of the truncation of the gas distribution, simultaneous
fitting of the background, and the absence of subjectively weighted 2D
temperatures \citep{Vikhlinin06b} are unique features of the JACO code
and are described in detail in \cite{Mahdavi07}.

The measured masses are robust to variations of the gas truncation
radius; varying this radius between $r_{200}$ and $r_{50}$ yields a
$3\%$ random variation in the measured masses. $T_{100}$ was often
poorly constrained (only lower limits were possible with lower quality
data). The JACO error analysis routines include this uncertainty in
the final masses.

The resulting total gravitating masses are listed in Table
\ref{tbl:xray}. In Figure \ref{fig:xraycomp} we compare these masses
with the values reported for the same Chandra observations in
\cite{LaRoque06} and \cite{Allen07}. We find that our masses on the
average in excellent agreement with the two samples, with the
\cite{LaRoque06} values being somewhat lower, and the \cite{Allen07}
values somewhat higher, than ours.

\begin{table*}
\begin{minipage}{166mm}
\caption{X-ray Measurements}
\label{tbl:xray}
\begin{tabular}{lrrcccccc}
\hline
Cluster & $N_\beta$ & $\chi^2/\nu$ & $M_{2500}$ & $M_{1000}$ & $M_{500}$ & $f_{2500}$ & $r_\mr{cool}$ & Extrapolated\\
\hline
    Abell   68 & 1 & 98/98    & $3.18 \pm  1.96$ & $ 4.98 \pm  4.10$ &$  6.63 \pm  6.15$ & $0.09 \pm  0.06$ &   $\ldots$        & 1000,500\\
    Abell  209 & 1 & 184/179  & $2.00 \pm  0.45$ & $ 3.97 \pm  1.77$ &$  5.91 \pm  2.98$ & $0.12 \pm  0.03$ &   $\ldots$        & 1000,500\\
    Abell  267 & 1 & 317/303  & $2.27 \pm  0.31$ & $ 4.08 \pm  0.98$ &$  6.26 \pm  2.05$ & $0.10 \pm  0.01$ & $0.030 \pm 0.030$ & 500\\
    Abell  370 & 1 & 915/804  & $2.99 \pm  0.24$ & $ 5.81 \pm  1.47$ &$  9.22 \pm  1.28$ & $0.12 \pm  0.01$ &   $\ldots$        & 500\\
    Abell  383 & 3 & 402/366  & $1.22 \pm  0.06$ & $ 3.04 \pm  0.65$ &$  5.03 \pm  1.34$ & $0.10 \pm  0.01$ & $0.114 \pm 0.014$ & 500\\
    Abell  963 & 3 & 1129/1071& $1.92 \pm  0.21$ & $ 3.24 \pm  0.52$ &$  4.20 \pm  1.64$ & $0.11 \pm  0.01$ & $0.079 \pm 0.008$ & 1000,500\\
    Abell 1689 & 3 & 1065/917 & $4.47 \pm  0.89$ & $ 7.40 \pm  2.32$ &$  9.86 \pm  3.23$ & $0.12 \pm  0.02$ & $0.082 \pm 0.015$ & 500\\
    Abell 1763 & 1 & 440/419  & $2.56 \pm  0.17$ & $ 4.78 \pm  0.50$ &$  7.68 \pm  1.11$ & $0.12 \pm  0.01$ &   $\ldots$        & 500\\
    Abell 2218 & 2 & 790/745  & $2.43 \pm  0.49$ & $ 2.89 \pm  0.89$ &$  5.99 \pm  2.33$ & $0.11 \pm  0.02$ &   $\ldots$        & 1000,500\\
    Abell 2219 & 1 & 2272/2249& $4.62 \pm  0.47$ & $10.88 \pm  3.18$ &$ 17.32 \pm  6.50$ & $0.11 \pm  0.01$ &   $\ldots$        & 1000,500\\
    Abell 2390 & 3 & 3663/3341& $3.75 \pm  0.31$ & $ 7.53 \pm  1.09$ &$ 10.87 \pm  2.09$ & $0.14 \pm  0.01$ & $0.099 \pm 0.009$ & 1000,500\\
CL 0024.0+1652 & 1 & 165/127  & $1.86 \pm  0.34$ & $ 4.87 \pm  0.74$ &$  9.92 \pm  9.90$ & $0.09 \pm  0.02$ & $0.073 \pm 0.023$ & 1000,500\\
MS 0015.9+1609 & 1 & 632/580  & $2.67 \pm  0.19$ & $ 6.52 \pm  1.09$ &$ 10.28 \pm  2.42$ & $0.21 \pm  0.01$ &   $\ldots$        & $\ldots$\\
MS 0906.5+1110 & 2 & 587/549  & $1.59 \pm  0.19$ & $ 2.49 \pm  0.45$ &$  3.23 \pm  0.73$ & $0.11 \pm  0.01$ & $0.030 \pm 0.030$ & 500\\
MS 1358.1+6245 & 2 & 752/675  & $2.83 \pm  0.78$ & $ 6.31 \pm  2.81$ &$ 11.08 \pm  6.33$ & $0.09 \pm  0.02$ & $0.077 \pm 0.012$ & 500\\
MS 1455.0+2232 & 3 & 1869/1685& $1.40 \pm  0.04$ & $ 2.25 \pm  0.10$ &$  2.96 \pm  0.16$ & $0.15 \pm  0.01$ & $0.123 \pm 0.010$ & $\ldots$\\
MS 1512.4+3647 & 1 & 254/243  & $0.69 \pm  0.13$ & $ 1.15 \pm  0.28$ &$  1.74 \pm  0.44$ & $0.12 \pm  0.02$ & $0.095 \pm 0.028$ & $\ldots$\\
MS 1621.5+2640 & 1 & 237/255  & $1.54 \pm  0.21$ & $ 3.61 \pm  0.54$ &$  5.43 \pm  1.22$ & $0.13 \pm  0.02$ &   $\ldots$        & $\ldots$\\
\hline
\end{tabular}

\medskip
All masses at density contrast $\Delta$ are in units of $10^{14}
M_\odot$ and measured within the lensing radius $r_\Delta$ as listed
in Table \protect\ref{tbl:lensing}.  $N_\beta$ is the number of
$\beta$-models required to achieve a good fit; $\chi^2/\nu$ is the
ratio of the $\chi^2$ statistic to the number of degrees of freedom
$\nu$ for the simultaneous fit to all spectra in all annuli; all
masses are in units of $10^{14} M_\odot$. The cumulative gas fraction
$f_{2500}$ is the ratio of the gas mass to the total mass at
$r_{2500}$; $r_\mr{cool}$ is the radius (in units of Mpc) within which
the isobaric cooling time is less than the age of the universe at the
redshift of the cluster. Also shown are the density contrasts at which
the masses had to be extrapolated beyond the available field-of-view.
\end{minipage}
\end{table*}

\begin{figure}
\begin{tabular}{c}
\includegraphics[width=76mm]{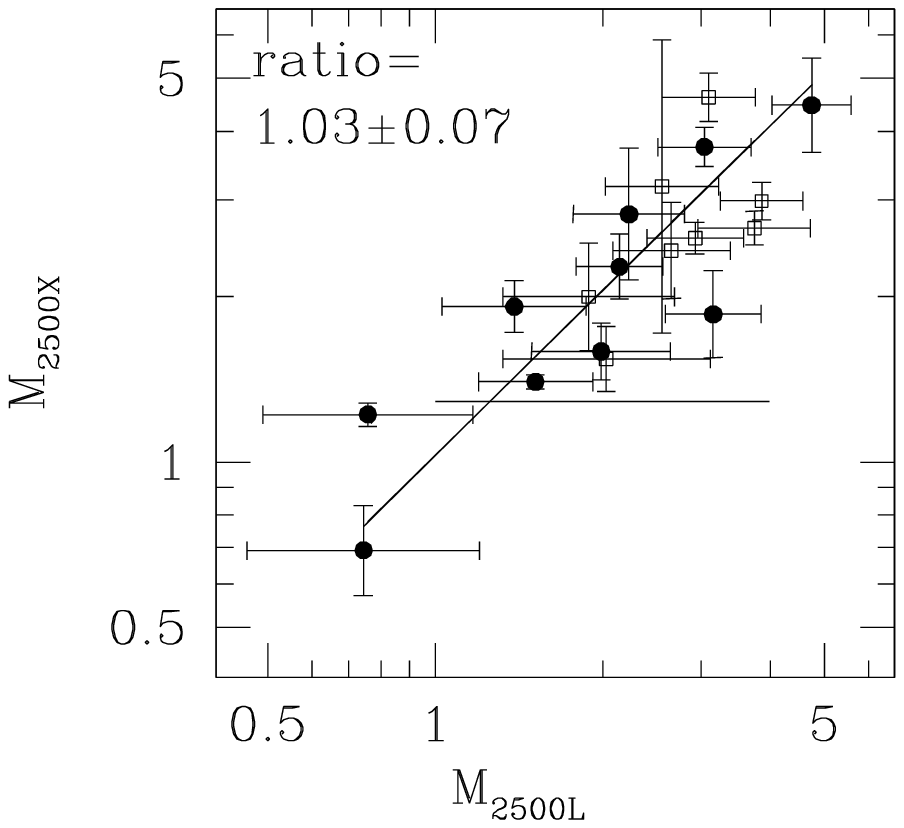} \\
\includegraphics[width=76mm]{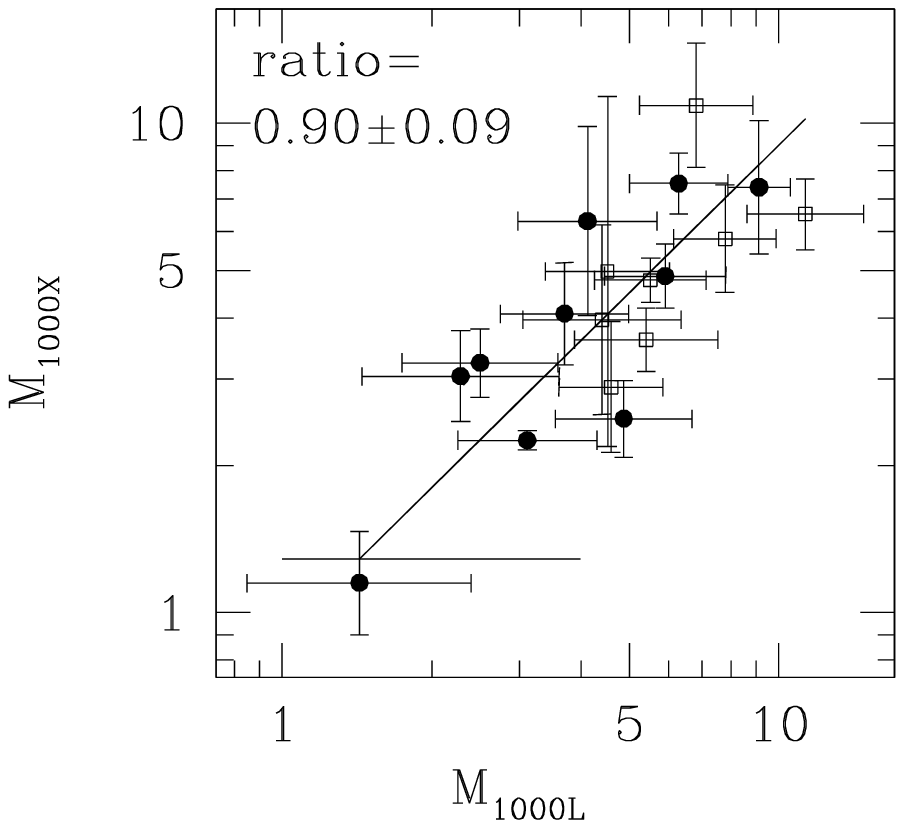} \\
\includegraphics[width=76mm]{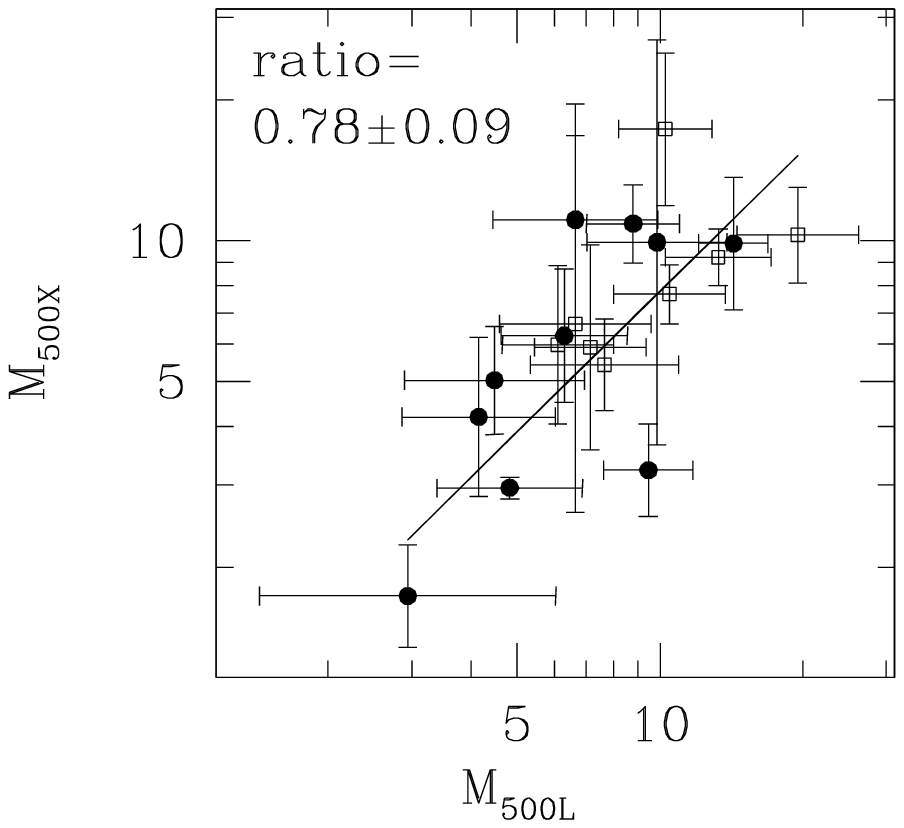} 
\end{tabular}
\caption{Ratio of the hydrostatic X-ray mass to the gravitational
lensing mass, both measured within the lensing overdensity radii
$r_{2500}$, $r_{1000}$, and $r_{500}$. All masses are in units of
$10^{14} M_\odot$. There is significant trend of decreasing $M_X/M_L$ with
increasing radius. Filled circles and unfilled squares show systems
with and without cool cores, respectively (we define non-cool core clusters
as those where the cooling time is nowhere smaller than the age of the universe
at the redshift of the cluster).}
\label{fig:results}
\end{figure}

\section{The $M_X/M_L$ Ratio}
\label{sec:trend}

\subsection{A Radial Trend}
\label{sec:fit}

We examine the relationship between the X-ray and lensing masses at
each density contrast using a simple constant of proportionality:
\begin{equation}
M_{\Delta,X} = a_\Delta M_{\Delta,L}
\end{equation}
Both $M_X$ and $M_L$ are measured within the weak-lensing derived
value of $r_\Delta$.  To estimate $a_\Delta$ properly, we minimize a
modified $\chi^2$ statistic appropriately weighted for errors in both
coordinates \citep{Press92}:
\begin{equation}
\chi^2 = \sum \frac{(M_{\Delta,X} - a_\Delta M_{\Delta,L})^2}
            {\sigma_{\Delta,X}^2 + a_\Delta^2 \sigma_{\Delta,L}^2}
\end{equation}
This formulation implies no intrinsic scatter in the data, an
assumption we validate below. We derive errors in $a_\Delta$ by
locating the values at which $\chi^2-\chi^2_\mr{min}=1$, which
correspond to the 68\% confidence interval.

The results appear in Figure \ref{fig:results}. The simple one-parameter
fit provides a good description of the relationship between 
$M_X$ and $M_L$ at all radii, and a clear trend is always
present. The goodness-of-fit figure $\chi^2/\nu = $ 16/17,
11/17, and 14/17 for masses measured within $r_{2500}$, $r_{1000}$,
and $r_{500}$, respectively. Therefore, there is no evidence of
intrinsic scatter in the $M_X - M_L$ relation. All scatter can be
explained by the statistical error. 

We find $a_{2500} = \rata$, $a_{1000} = \ratb$, and $a_{500} = \ratc$.
The formal significance of the result is higher than it seems from the
quoted errors, because the three slopes $a_\Delta$ are correlated and
therefore cannot be compared directly as a function of $\Delta$. This
is because the masses at each radius are correlated with the masses at
other radii. Wherever the X-ray masses are extrapolated (as shown in
Table \ref{tbl:xray}), they are highly correlated with $M_{2500,X}$;
the lensing masses at all radii are highly correlated, because the
aperture masses are derived by integrating over the tangential shear
from the radius of interest out to large radius.

\begin{figure}
\includegraphics[width=84mm]{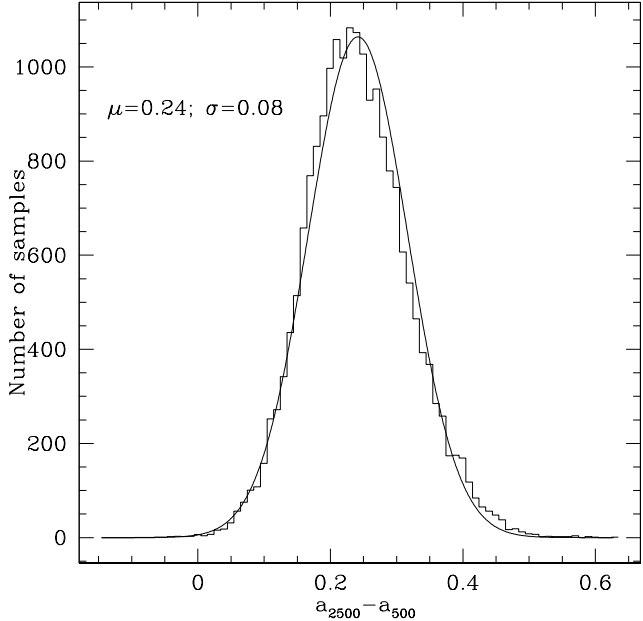}
\caption{The histogram shows the probability distribution of the
  difference between $M_x/M_L$ measured at $r_{2500}$ and $M_x/M_L$
  measured at $r_{500}$.  The difference is positive at $3\sigma$
  significance. For comparison, we plot a normal distribution
 with $\mu=0.24$, $\sigma=0.08$.}
\label{fig:bootstrap}
\end{figure}

To properly evaluate the significance of the difference between
$a_{2500}$ and $a_{500}$, we employ a bootstrap procedure with,
calculating $a_{2500}-a_{500}$ many times. The resulting distribution
appears in Figure \ref{fig:bootstrap}. We find that $a_{2500}-a_{500}$
follows a Gaussian distribution with a mean of 0.24 and a standard
deviation of 0.08. Thus, the difference in the $M_X/M_L$ ratios at
$r_{2500}$ and $r_{500}$ is significant at the 3$\sigma$ level. Using
a Monte Carlo simulation of the bootstrap processes, we confirm that
the error obtained in this way is unbiased and the ``error on the
error'' is close to the expected value, $0.07/\sqrt{36} \approx 0.01$,
or 16\%.

\subsection{Interpretation}

\begin{figure}
\includegraphics[width=84mm]{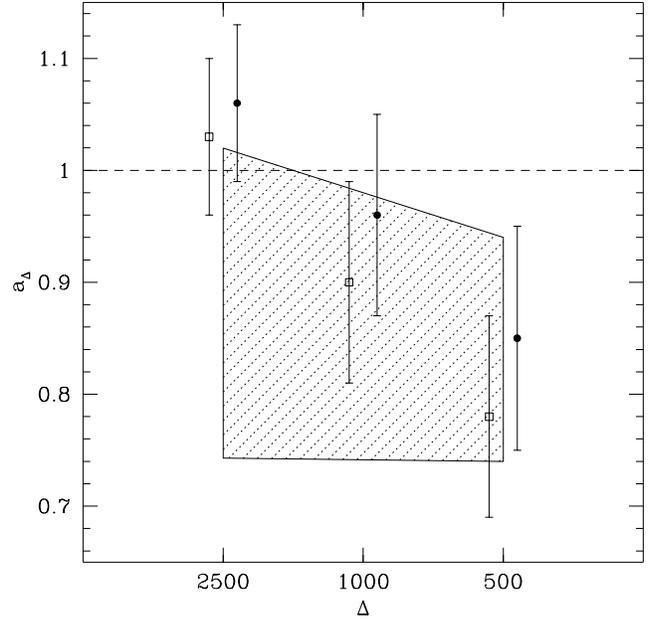}
\caption{The $M_X/M_L$ ratio as a function of overdensity $\Delta.$
  The unfilled squares show the data
  before correction for excess structure along the line of
  sight as described by \protect\cite{Johnston07}; the filled 
  circles show the data after correction for the effect.
  The trend is consistent with the X-ray mass underestimate
  predicted by N-body work \protect\citep[shaded region, ][]{Nagai07}.}
\label{fig:twohalo}
\end{figure}

The variation of this constant of proportionality $a_\Delta$ as a
function of radius can yield useful constraints on cluster
physics. This statistically significant gradient in $M_X/M_L$ may be
explained through an X-ray mass underestimate. For example, the
cosmological N-body simulations discussed by \cite{Nagai07} predict a
low $M_X/M_\mr{true}$ on account of nonthermal pressure support in the
ICM: effects such as residual bulk motions cause hydrostatic mass
estimates to be biased low.  Earlier studies based on simulations
\citep[e.g.][]{Evrard90,Faltenbacher05,Rasia06,Hallman06} support this
conclusion. \cite{Nagai07} find that, on the average, clusters at
$z=0$ exhibit little significant bias within $r_{2500}$ ($-0.12 \pm
0.16$). However, within $r_{500}$, the fractional bias is more
significant ($-0.16 \pm 0.10$). Assuming that weak gravitational
lensing mass has been corrected for projection effects (see below),
the direction and magnitude of the effect predicted by \cite{Nagai07}
at $r_{500}$ would seem to match our observed trend (see Figure
\ref{fig:twohalo}).

An independent check on this result is a comparison with previously
reported normalization of the $M_L - T$ relation. At $r_{2500}$,
\cite{Hoekstra07} finds consistency between the normalizations of the
$M_L - T$ from CFHT data and the $M_X - T$ relation from
\cite{Vikhlinin06}. On the other hand, at $r_{500}$ \cite{Pedersen07}
finds the normalization to be high by $\approx 30\%$. These results
are consistent with the trend we detect here.

If, as we suggest, incomplete thermalization of ICM is responsible for
the X-ray mass underestimate, we should expect clusters with the most
substructure to exhibit the greatest deviations from $M_X = M_L$
\citep[e.g. Abell 781,][]{Sehgal07}.  Recent theoretical work has
shown that it may be possible to correct for the X-ray mass
underestimate in this way \citep{Jeltema07}. We look for such a trend
in Figure \ref{fig:results} by dividing the clusters into cool core
and non-cool core subsamples, under the hypothesis that clusters with
cool cores are less likely to have experienced recent mergers. We find
that there is no statistically significant difference in the $M_X -
M_L$ relation for cool core and non-cool core clusters. This agrees
with N-body simulations showing that the existence of a cool core is
not a reliable predictor of the equilibrium state of a cluster of
galaxies \citep[e.g.][]{Poole06,Poole07,Burns07}. More refined estimates of the level
of substructure in each cluster are needed to evaluate the possibility
of correcting for the X-ray underestimate. We will explore this avenue
in a future paper.

\section{Potential Biases}
\label{sec:bias}

As discussed above, we interpret the change in $M_X/M_L$ with
overdensity $\Delta$ as evidence of a deviation from hydrostatic
equilibrium. However, an obvious concern is whether the result can be
explained by biases in our measurements. After all, the geometries of
real clusters are more complicated than assumed here (although the
ensemble averaged system should be close to spherical). In this
section we focus on a number of potential effects that could lead to a
radial dependence. Note that other effects can change $M_X/M_L$, but do
so independent of scale; for instance a change in the mean source
redshift simply changes all $M_\Delta$'s by the same factor.

\subsection{Elongation Bias}

An explanation of our results could be that we are introducing a
systematic bias through spherical modeling of what we know ought to be
triaxial systems. \cite{Piffaretti03} consider this question in
detail. They find that for a broad distribution of elongations and
inclinations, the differences between triaxial and spherical X-ray
mass estimates are negligible, of order 3\%. This fraction is much
smaller than the effect we observe. For clusters elongated nearly
exactly along the line of sight, \cite{Piffaretti03} do find that
assuming spherical symmetry causes the projected X-ray mass to be
underestimated by up to 30\% at $r_{500}$.  It is conceivable that
through X-ray selection, a number of clusters in our samples might be
preferentially elongated close to the line of sight.  However, even if
present, such elongations cannot reproduce the trend we observe. As
\cite{Piffaretti03} show, geometries that yield a 30\% projected mass
deficit at $r_{500}$ ought to show a similar deficit at $r_{2500}$,
which we do not observe. Thus while the overall normalization of the
trend in Figure \ref{fig:twohalo} might be affected by preferential
selection of elongated objects, the trend itself cannot be.

\subsection{Extrapolation Bias}

We also investigate whether the fact that most $M_{500,X}$ are
extrapolated could affect the results. The four clusters for which no
X-ray mass extrapolations are necessary are MS 0016, MS 1455, MS 1512,
and MS 1621. For these clusters alone, we find $a_{2500} = 0.83 \pm
0.13$, and $a_{500} = 0.61 \pm 0.13$.  The difference in these ratios
is similar to that of the full sample. Thus there is no indication
that the substantial difference between $a_{2500}$ and $a_{500}$ is
caused by the extrapolation of the X-ray masses. For the extrapolated
sample only, we obtain $a_{2500} = 1.07 \pm 0.08$ and $a_{500} = 0.84
\pm 0.11$. Thus the overall ratios for the extrapolated and
unextrapolated subsamples do not differ significantly (the subsamples
are uncorrelated, and thus at fixed $\Delta$ we can fairly compare the
nominal errors in $a_\Delta$ for the extrapolated and unextrapolated
subsamples).

\subsection{Dependence on the Mass-Concentration Relation}

We check whether our use of the \cite{Bullock01} mass-concentration
relation to deproject the weak lensing masses could cause the observed
trend. We vary the normalization and redshift dependence of the
mass-concentration relation. The corresponding mean concentration
varies by $\sim 30\%$ (which we note is a rather extreme change),
which leads to a systematic variation in the $M_{2500,L}/M_{500,L}$
ratio less than 5\%. More recent N-body work shows that the slope of
the mass-concentration relation may be somewhat shallower than the
\cite{Bullock01} value \citep[e.g.][]{Neto07}. However, because our
sample spans only an order of magnitude in mass, the effect of
changing the slope to the \cite{Neto07} value (-0.1) is negligible. We
therefore conclude that the use of an incorrect NFW concentration
cannot reproduce a result of the strength we observe.

\subsection{Two-Halo Term Correction}

Finally, we consider the possibility that deprojections of weak
lensing mass measurements \emph{in general} are biased.  When
considering the accuracy of weak lensing masses at better than the
20\% level, an important effect is the contribution of nearby large
scale structure to the weak lensing mass\footnote{ As opposed to
  distant large scale structure which is a source of random noise and
  not a bias \citep{Hoekstra01}.}.  The recent detection of a
``two-halo term''---excess mass due to halos and uncollapsed material
outside the virial radius---in Sloan Digital Sky Survey (SDSS)
clusters \citep{Johnston07} provides direct observational evidence for
considering such a contribution.

The ``two-halo'' term could be an important effect in our
measurements. We have carried out the deprojection of the weak lensing
masses assuming that they are described by an NFW profile. However, in
\cite{Johnston07} SDSS data, shear measurements out to 30 Mpc indicate
that outside the virial radius, the matter density profile gradually
flattens. For the range of masses we consider, the \cite{Johnston07}
``two-halo'' density profile is very approximately represented by the
form
\begin{equation}
\rho_{2H} \propto \frac{1+(0.25 r/r_{200})^2}{r (r_{200} + c r)^2}
\label{eq:2h}
\end{equation}
The above equation resembles an NFW profile within $r_{200}$, but
acquires a flatter $r^{-1}$ shape outside $4 r_{200}$. The profile
is truncated at 30 Mpc, near where the cluster-mass correlation
function becomes negligible \citep{Hayashi07}.

If equation \ref{eq:2h} describes the correct density profile, and yet
we mistakenly use an NFW profile to deproject the lensing mass, then
our 3D mass estimates would exceed the true value. It is
straightforward to calculate exactly how much we would overestimate
the masses; we can project both the NFW and the ``two-halo'' profile
along the line of sight, and take the ratio of the estimated NFW and
``two-halo'' 3D masses. This ratio then represents a correction that we
can apply to our data on a case-by-case basis.  The results appear in
Figure \ref{fig:twohalo}. We find that while the two-halo term
correction does not remove the trend in $a_\Delta$, it does mitigate
it, particularly with regards to the large deficit seen at $r_{500}$.

The $M_X/M_L$ trend remains because any extraneous mass along the line
of sight affects mass estimates at all radii, and the difference
between the excess at $r_{2500}$ and $r_{500}$ is not sufficient to
remove the radial trend in the uncorrected data. The corrected values
are $a_{2500} = \twrata$, $a_{1000} = \twratb$, and $a_{500} =
\twratc$. If we repeat the bootstrap procedure in Section
\ref{sec:fit}, we find that the difference between $a_{2500}$ and
$a_{500}$ is reduced to $0.20 \pm 0.08$, still significant at
$2.5\sigma$, i.e. better than the 98\% confidence level.

\section{Gas Fraction Trends}
\label{sec:fgas}

We also consider the X-ray gas fraction $f_\mr{gas}$ as a function of
lensing mass. The gas fraction is an important quantity for cosmology
with clusters of galaxies.  Under the assumption that $f_\mr{gas}$
approaches a universal value at large enough radii, it is possible to
derive constraints on the dark energy equation of state
\citep[see][and references therein]{Allen07}. The constraints depend
crucially on the assumption of universality; if $f_\mr{gas}$ were not
a universal number independent of mass and redshift, the dark energy
constraints would not be correct \citep[e.g.][]{Vikhlinin06}.

\begin{figure}
\begin{tabular}{c}
\includegraphics[width=84mm]{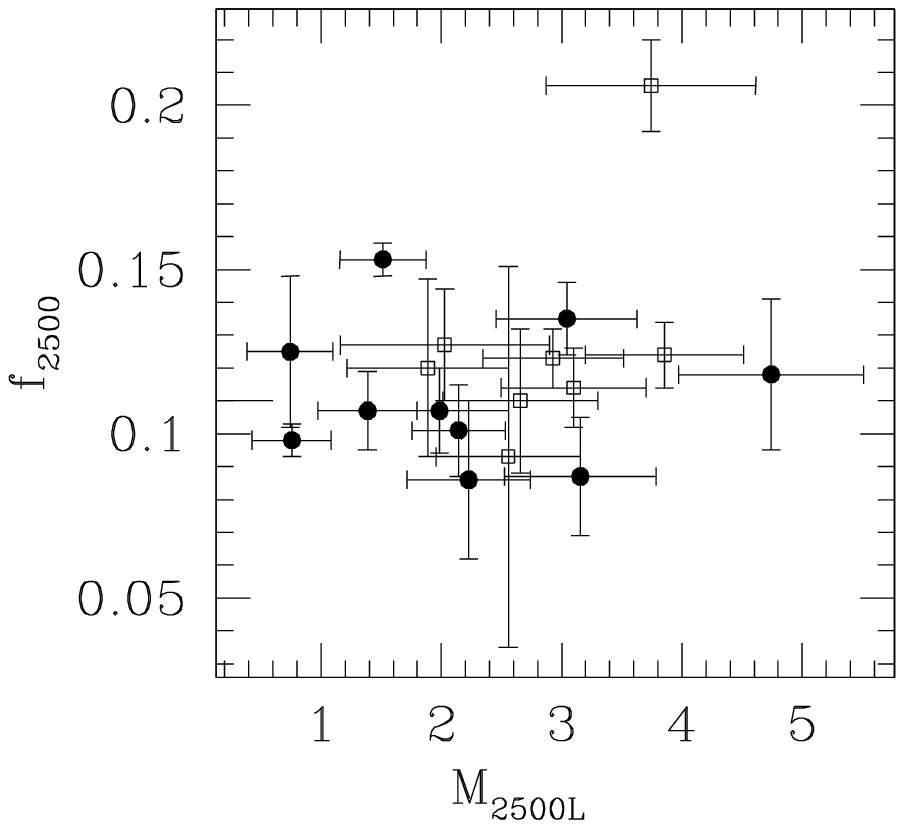}\\
\includegraphics[width=84mm]{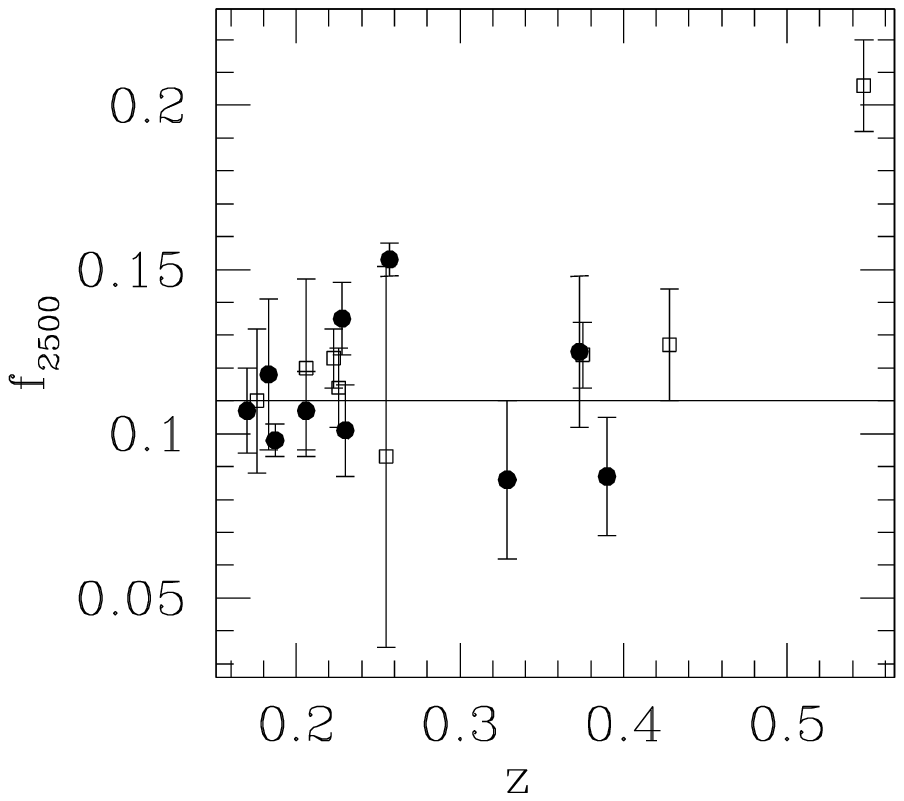}
\end{tabular}
\caption{The X-ray gas fraction as a function of (\emph{top}) weak
  lensing mass at $r_{2500}$ and (\emph{bottom}) redshift. There is no
  evidence of a trend with mass or redshift. Filled circles and unfilled squares
  show systems with and without cool cores, respectively. The solid 
  horizontal line shows the \protect\cite{Allen07} value.}
\label{fig:fgas}
\end{figure}

In Figure \ref{fig:fgas} we show the X-ray gas fraction as a function
of the lensing mass at $r_{2500}$. There is remarkably little scatter
in the relation. The clearest outlier in $f_\mr{gas}$ is MS 0015.9, a
previously known high $f_\mr{gas}$ system with triaxial structure
elongated along the line of sight \citep{Piffaretti03}. Otherwise, the
gas fractions are closely clustered around $0.1$. The mean value for
all systems is $\langle f_{2500}\rangle = 0.119 \pm 0.006$. This value
is in excellent agreement with the value found by \cite{Allen07} for
42 clusters, $f_{2500} = 0.110 \pm 0.002$. 

Importantly, $f_{2500}$ is uncorrelated with both $M_{2500,L}$ and $z$
(Kendall's $\tau$ is 0.079 and 0.217, respectively).  We therefore
cannot reject the hypothesis that $f_{2500}$ is independent of lensing
mass and redshift.

\section{Conclusion}
\label{sec:conclusion}

We provide a comparison of hydrostatic X-ray and weak lensing masses
for a sample of 18 galaxy clusters. What makes our analysis unique is
the uniform analysis of Chandra X-ray data and CFHT weak lensing data.
This significantly improves the reliability of the comparison.  At
$r_{2500}$ we find excellent agreement between lensing and X-ray
masses: we obtain a ratio $M_X/M_L = \rata$. Interestingly, we observe
a significant decrease in this ratio towards larger radii.  At
$r_{500}$ the ratio is $M_X/M_L = \ratc$. Accounting for correlations
between the mass measurements, we find that the difference in the
ratios is $0.24\pm0.08$, significant at $3\sigma$. The trend of
$M_X/M_L$ with radius is consistent with previous measurements of the
$M_L - T$ relation, and is not caused by the assumption of spherical
symmetry, extrapolation beyond the Chandra field of view, or
uncertainty in the mass-concentration relation.

The trend remains even after we correct for a systematic overestimate
of the weak lensing mass due to correlated large scale structure.  We
show that the trend is consistent with simulations in which nonthermal
pressure support causes a systematic underestimate of the cluster mass.
Interestingly, the underestimate is not correlated with the presence
or absence of a cool core.  These results are relevant for upcoming
surveys that aim to measure the cosmological parameters from large
cluster samples, especially for observations seeking measurements
closer to $r_{500}$.

We also determine the X-ray gas fraction at $r_{2500}$; this quantity
has been used in several studies to measure $w$, the parameter of the
dark energy equation of state. We find that $f_{2500}$ is not
correlated with lensing mass or redshift.

We thank Gus Evrard and  Andrey Kravtsov for enlightening
discussions. AB and HH acknowledge support from NSERC. AB also
acknowledges support from the Leverhulme trust in the form of a
visiting professorship at Oxford and Durham Universities. HH also
acknowledges support from the Canadian Insititute for Advanced
Research and grants from NSERC, CFI and BCKDF. This work was partially
supported by NASA grant NNX07AE73G.  Additional research funding was
provided by J. Criswick.

\ifthenelse{\equal{\mode}{submit}}{

}{\bibliographystyle{myrefs/mn2e}
\bibliography{myrefs/myrefs} }
\end{document}